\documentclass[aps,prl,showpacs,twocolumn,groupedaddress]{revtex4-1}

\usepackage{graphicx,graphics,color,epsfig}
\usepackage{amsmath}
\usepackage{amssymb}
\usepackage{amsfonts}
\usepackage{amsfonts}
\usepackage{epstopdf}
\usepackage{bm}
\usepackage{times,xspace}
\usepackage{color}

\begin{document}


\title{Spin-fluctuations and the peak-dip-hump feature in the photoemission spectrum of actinides}

\author{Tanmoy Das}
\affiliation{ Los Alamos National Laboratory, Los Alamos, New Mexico 87545 USA}

\author{Jian-Xin~Zhu}
\affiliation{ Los Alamos National Laboratory, Los Alamos, New Mexico 87545 USA}

\author{Matthias J.~Graf}
\affiliation{ Los Alamos National Laboratory, Los Alamos, New Mexico 87545 USA}

\date{\today}

\begin{abstract}
We present first-principles multiband spin susceptibility calculations within the random-phase approximation
for four isostructural superconducting PuCoIn$_5$, PuCoGa$_5$, PuRhGa$_5$, and nonsuperconducting UCoGa$_5$ actinides. The results show that a strong peak in the spin-fluctuation dressed self-energy is present around 0.5 eV in all materials,
which is mostly created by 5$f$ electrons. These fluctuations couple to the single-particle spectrum and give rise to a peak-dip-hump feature, characteristic of the coexistence of itinerant and localized electronic states. Results are in quantitative agreement with photoemission spectra. Finally, we show that the studied actinides can be understood within the rigid-band filling approach, in which the spin-fluctuation coupling constant follows the same materials dependence as the superconducting transition temperature $T_c$.
\end{abstract}

\maketitle

The discovery of superconductivity in PuCoGa$_5$~\cite{SarraoYEAR02} and soon thereafter in isostructural PuRhGa$_5$~\cite{FWastin:2003}, and PuCoIn$_5$~\cite{EDBauer:2011} (collectively called Pu-115 series) has revitalized the interest in  the spin-fluctuation mechanism of high-temperature superconductivity. In particular, a systematic study of
spin-fluctuation temperature $T_s$ versus superconducting transition temperature $T_c$ indicates that Pu-115 compounds lie in between the Ce-based 4$f$-electron heavy-fermion and $d$-electron superconductors (cuprates and pnictides) \cite{CurroYEAR05}. Within the actinide series, the duality of correlation effects in plutonium compounds
stems from Pu's position between the itinerant 5$f$ states of uranium \cite{SmithU} and the localized 5$f$ states of americium~\cite{SmithA}. This makes Pu a unique candidate to define the intermediate coupling regime of Coulomb interaction in which neither the purely itinerant mean-field theory nor the strong-coupling Kondo lattice model
hold exactly --- a prototypical example of strongly correlated electron systems~\cite{QMSi:2009}. On the other hand, the diagrammatic perturbation theory of fluctuations can still be applied as long as the Hubbard $U\sim W$, where $W$ is the non-interacting bandwidth \cite{Dasopt}. Therefore, it is important to characterize the evolution of the spin-fluctuation excitations in Pu-115 compounds, which will help to delineate the role of spin-fluctuation mediated superconductivity in $f$-electron systems.

Photoemission spectroscopy (PES) has revealed a strong spectral weight redistribution in the single-particle spectrum with a prominent peak-dip-hump structure around 0.5 eV in PuCoGa$_5$~\cite{JoycePRL03}. This feature has been interpreted as the separation between itinerant (peak) and localized (hump) electronic states of the 5$f$ electrons~\cite{MEPezzoli:2011,JXZhu:2011}.
To provide insights into this PES structure, we present a first-principles multiband spin susceptibility calculation within the random-phase approximation (RPA). The results show that a considerably large amount of the spin-fluctuation instability is present in the 0.5~eV energy range which originates from the particle-hole channel between 5$f$ states. The resulting self-energy correction due to spin fluctuations is calculated within the GW approach, which quantitatively reproduces the observed peak-dip-hump PES feature in PuCoGa$_5$.

We interpret the spin-fluctuation effects on PES along the same line as the localized-itinerant
duality discussed above. The fluctuation spectrum creates a dip in the single-particle excitations due to strong scattering. The lost spectral weight (dip) is distributed partially to the renormalized itinerant states at the Fermi level (peak), as well as to the strongly localized incoherent states at higher energy (hump). The coherent states at the Fermi level can still be characterized as Bloch waves, though renormalized, whereas the incoherent electrons are localized in real space exhibiting the dispersionless hump feature. We perform calculations for the actinide materials PuCoIn$_5$ ($T_c=2.4$~K), PuCoGa$_5$ ($T_c=18.5$~K), PuRhGa$_5$  ($T_c=8.7$~K),  and UCoGa$_5$ ($T_c=0$~K), which show that the computed spin fluctuations play a significant role for the systematic evolution of the electronic band renormalization and spectral weight redistribution across these compounds. 
{We also deduce the computed spin-fluctuation coupling constant $\lambda$, which follows $T_c$ as we move across the series from PuCoIn$_5$$\rightarrow$PuCoGa$_5$$\rightarrow$PuRhGa$_5$$\rightarrow$UCoGa$_5$, suggesting that spin fluctuations play a crucial role in the pairing mechanism.} The results also demonstrate that the actinides can be understood within a unified description of rigid-band shift of the 5$f$ electrons close to the Fermi level (hole doping).

{\it Intermediate coupling model$-$} We calculate materials specific first-principles electronic band structures,
including spin-orbit coupling, within the framework of density functional theory in the generalized-gradient
approximation (GGA)~\cite{JPPerdew:1996}. We use the full-potential linearized augmented plane wave method of WIEN2k~\cite{PBlaha:2001}. The calculation is performed with 40 bands to capture the $\pm$10 eV energy window of  relevance around the Fermi level. The non-interacting susceptibility in the particle-hole channel represents joint density of states (JDOS), which can be calculated by convoluting the multiorbital Green's function $G_{sp}({\bf k},i\omega_n)$ ($s,p$ are orbital indices), to obtain (spin and charge bare susceptibility are the same in the paramagnetic ground state)\cite{Graser}:
\begin{eqnarray}\label{chibare}
\chi^0_{spqr}({\bm q},\Omega)=-\frac{T}{N}\sum_{{\bm k},n}G_{sp}({\bm k},i\omega_n)G_{qr}({\bm k}+{\bm q},i\omega_n+\Omega).
\end{eqnarray}
Within the RPA, spin and charge channels become decoupled. (We ignore particle-particle as well as  weaker charge fluctuation processes.) In the spin-channel, the collective many-body corrections of the spin fluctuation spectrum can be written in matrix representation: $\hat{\chi}=\hat{\chi}^0[\hat{1}-\hat{U}_s\hat{\chi}^0]^{-1}$. The interaction matrix $\hat{U}_s$ is defined in the same basis consisting of intra-orbital $U$, inter-orbital $V$, Hund's coupling $J$ and pair-scattering $J^{\prime}$ terms.\cite{Graser,Dastworesonance,DasFeSe} In the present calculation, we neglect the orbital overlap of eigenstates, and hence $\hat{\chi}^0$ becomes a diagonal matrix and $J=J^{\prime}=0$.

Using the GW approximation, where $G$ represents the Green's function and $W$ is the interaction vertex, we write the spin-fluctuation interaction vertex  following Ref.~\cite{Bickers} as $V_{pqrs}({\bf q},\Omega) = \left[\frac{3}{2}\hat{U}_s \hat{\chi}^{\prime\prime}({\bm q},\Omega)\hat{U}_s\right]_{pqrs}.$ The Feynmann-Dyson equation for the  imaginary part of the self-energy in a multiband system with $N$  sites at $T=0$ is (for details see the supplement \cite{supplement})
\begin{eqnarray}
\Sigma_{pq}^{\prime\prime}(\omega) = -2\sum_{rs}\int_{0}^{\omega} d\Omega\,
\Gamma \bigl<V_{pqrs}(\Omega)\bigr>_{\bm q} N_{rs}(\omega-\Omega),\label{Eq:selfenergy2}
\end{eqnarray}
for $\omega>0$, where the density of states is given by $N_{rs}(\epsilon)=-\sum_{\bm k}{\rm Im}\bigl[G_{rs}({\bm k},\epsilon)\bigr]/\pi$. (For $\omega < 0$, the only changes are that the upper limit of the integral is $|\omega|$ and the argument of $N_{rs}$ is $\Omega-|\omega|$, which is $<0$.) $\Gamma$ is the vertex correction discussed later. For a more accurate calculation, one needs to account for the anisotropy in $\hat V({\bf q},\Omega)$.
In the present case, where the spin-fluctuation spectrum is considerably isotropic (see Fig.~\ref{spin_1imp}), it is justified to use a momentum-averaged spin-fluctuation function.

We use Eq.~(2) to compute the imaginary part of the self-energy from the first-principles band structure. {The real part of the self-energy, $\Sigma_{pq}^{\prime}(\omega)$, is obtained by using the Kramers-Kronig relationship.} Finally, the self-energy dressed quasiparticle spectrum is determined by Dyson's equation: $\hat{G}^{-1}=\hat{G}_{0}^{-1}-\hat{\Sigma}$. The full self-consistency in the GW approximation requires the  dressed Green's function $\hat{G}$ to be used in $\hat{\chi}^0$. This procedure is numerically expensive, especially for multiband systems. To overcome this burden, we adopt a modified self-consistency scheme, where we expand the real part of the self-energy $\Sigma_{pq}^{\prime}\approx(1-Z^{-1})\omega$ in the low-energy region where $\Sigma_{pq}^{\prime\prime}\rightarrow 0$. The resulting self-energy dressed Green's function is used in Eqs.~(1)-(2) which keeps the formalism unchanged with respect to the renormalized band ${\bar \xi}^{\nu}_{\bm k}=Z\xi_{\bm k}^{\nu}$. In this approximation the vertex correction in Eq.~(2) simplifies to $\Gamma=1/Z$ according to the Ward identity. We note that all calculations are performed by solving matrix equations, while the results shown below are for the trace of each quantity. For brevity, we drop the symbol `trace' altogether.

\begin{figure}
\rotatebox[origin=c]{0}{\includegraphics[width=.95\columnwidth]{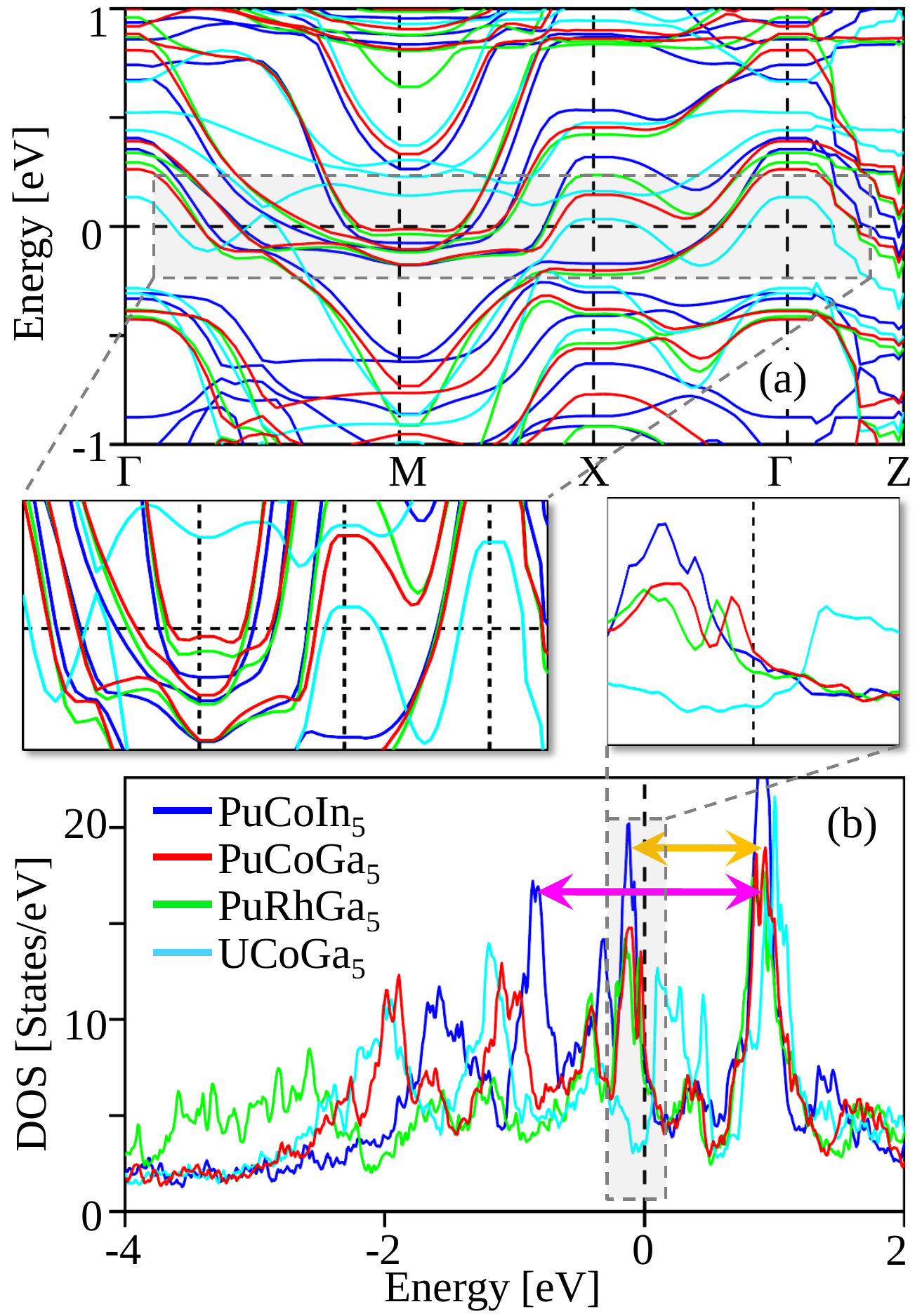}}
 \caption{(Color online) (a) First-principles GGA electronic band-structure calculations for various Pu-115 and UCoGa$_5$ actinides near $E_F$. (b) Corresponding DOS in the low-energy region of present interest. The arrows mark the relevant particle-hole excitations. {\it Insets:} Low-energy regions of dispersion and DOS showing that all materials are related by a rigid shift of bands in this energy scale.} \label{rho_1imp}
\end{figure}

{\it Results$-$} Figure~\ref{rho_1imp} presents the calculated GGA band structure in (a) and corresponding non-interacting DOS in (b) for all four materials studied here. Notice that the low-energy band structure remains very much the same for all materials. It only shifts upward in energy in moving along the series PuCoIn$_5$$\rightarrow$PuCoGa$_5$$\rightarrow$PuRhGa$_5$$\rightarrow$UCoGa$_5$. This behavior can be accounted for by a rigid band shift, see {\it insets} to Fig.~1. The Pu-115 compounds show two sharp peaks in the DOS just below and above $E_F$, which are mainly originated from the 5$f$ electrons of Pu atoms. The $3d$ (or $4d$) and $4p$ (or $5p$)
electrons of the reservoir elements are not important in this energy scale [See Refs.~\onlinecite{Saadthesis,JXZhu:2011} for partial DOS]. As the DOS at $E_F$ decreases in going to UCoGa$_5$  (see cyan lines in Fig.~\ref{rho_1imp}), most of the 5$f$ states move above $E_F$, reducing the correlation strength to a large extent.

\begin{figure}
\rotatebox[origin=c]{0}{\includegraphics[width=.95\columnwidth]{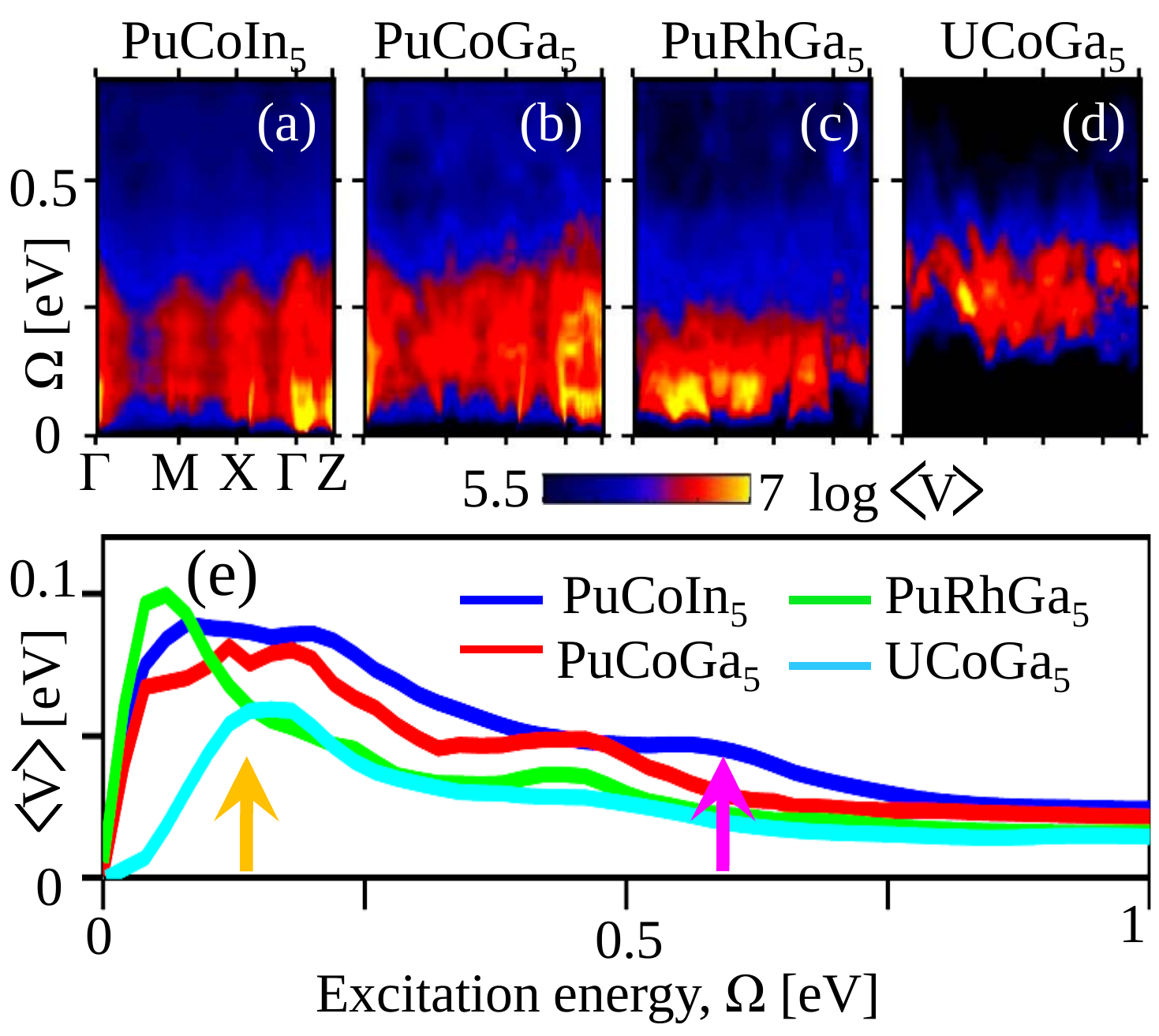}}
 \caption{(Color online) The spin-fluctuation vertex $V({\bf q},\Omega)$ is plotted along high-symmetry directions in (a)-(d). Panel (e): The corresponding $\bigl<V(\Omega)\bigr>_q$ averaged over 3D momentum space. All the calculations are performed for $\pm$10~eV energy window, but  results are shown only in the relevant energy region. }
 \label{spin_1imp}
\end{figure}

Projections of the computed spin-fluctuation vertex, $\hat V({\bf q},\omega)$, are plotted in Fig.~\ref{spin_1imp}  as a function of excitation energy along the high-symmetry momentum directions. Our choice of the screened Coulomb term $U$ satisfies the intermediate coupling approximation of $U/W\sim1$. As seen from the band structures in Fig.~1(a), the average bare bandwidth for all materials near the Fermi level is of order of 1~eV. Hence, we set $U$=1~eV for all compounds, which is below the critical value of a magnetic instability, that is, $U\chi^0(\bm{q},\omega=0)<1$ for all $\bm{q}$. Note that our screened $U$ for the spin-fluctuation calculation is smaller than that used in LDA+$U$ type calculations, where a rather large value of U~3 eV was introduced into the local orbital basis\cite{Pourovskii,phonon,LDMA}.

All spectra split mainly into two energy scales (at higher energy, no other prominent peak is seen in the computed spectra up to 10~eV and thus not shown). Corresponding momentum-averaged values $ \big<V\bigr>_{\bm q}$ are fairly similar for all Pu-115 compounds, but notable different for UCoGa$_5$. The low-energy peak arises from the transition between the 5$f$ states just below to above $E_F$ (within the RPA, the peak shifts to lower-energy), see gold arrow in Figs.~1(b) and 2(e). The high energy hump comes mostly from the transition of the second peak in the DOS below $E_F$ (hybridized $d$- and $p$-states also contribute~\cite{Saadthesis}) to the 5$f$ states above $E_F$ as marked by the magenta arrow in Figs.~1(b) and 2(e). For UCoGa$_5$ most of the 5$f$ states shift above $E_F$ and thus intra-orbital spin fluctuations do not survive, while the inter-orbital spin fluctuations move to higher energy.

\begin{figure}
\rotatebox[origin=c]{0}{\includegraphics[width=0.95\columnwidth]{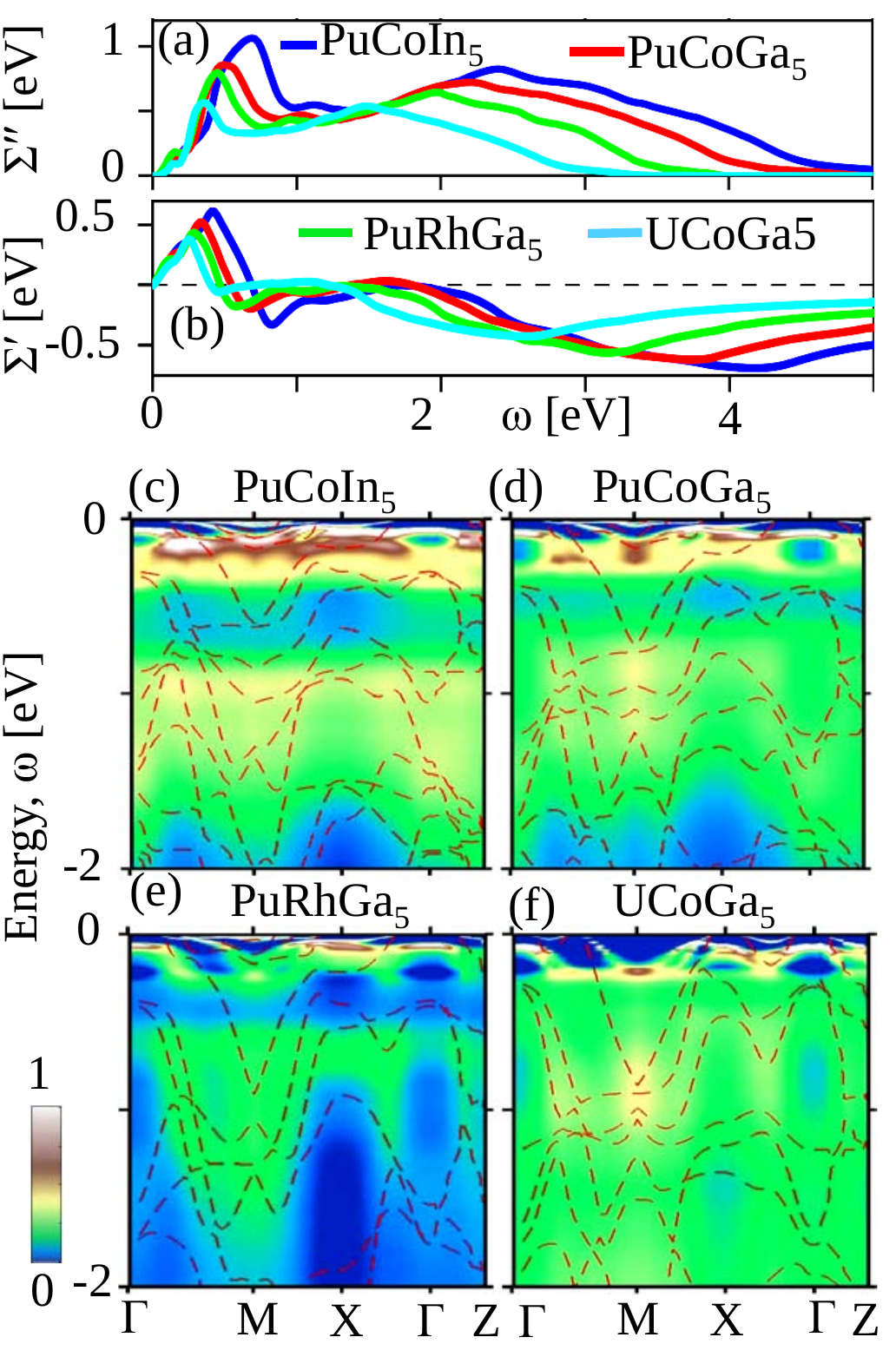}}
 \caption{(Color online) The computed momentum-averaged $\Sigma^{\prime\prime}(\omega)$ and $\Sigma^{\prime}(\omega)$ are plotted in (a) and (b), respectively. All  peak positions in $\bigl<V\bigr>$ in Fig.~\ref{spin_1imp}(b) are shifted to higher energy in $\Sigma^{\prime\prime}$ due to band-structure effects. Panels (c)-(f): The self-energy dressed angle-resolved spectral weight $A({\bf k},\omega)$ is plotted along high-symmetry momentum directions. The peak-dip-hump feature is clearly evident in all spectra below 1~eV.}\label{sig_1imp}
\end{figure}

The coupling of the spin fluctuations to the quasiparticle excitations gives the self-energy correction in Eq.~2.
The imaginary and real part of $\Sigma$ are plotted in Figs.~\ref{sig_1imp}(a) and \ref{sig_1imp}(b), respectively. Note that $\Sigma^{\prime\prime}(\omega)$ shows a peak-dip-hump feature, although strongly enhanced  by the DOS in comparison with $\bigl<V(\Omega)\bigr>_q$. Both the low- and high-energy features move toward $\omega=0$ as the 5$f$ states shift toward $E_F$ across the series PuCoIn$_5$$\rightarrow$PuCoGa$_5$$\rightarrow$PuRhGa$_5$$\rightarrow$UCoGa$_5$ (for UCoGa$_5$ the 5$f$ states eventually cross above $E_F$). 

\begin{figure}
\rotatebox[origin=c]{0}{\includegraphics[width=.9\columnwidth]{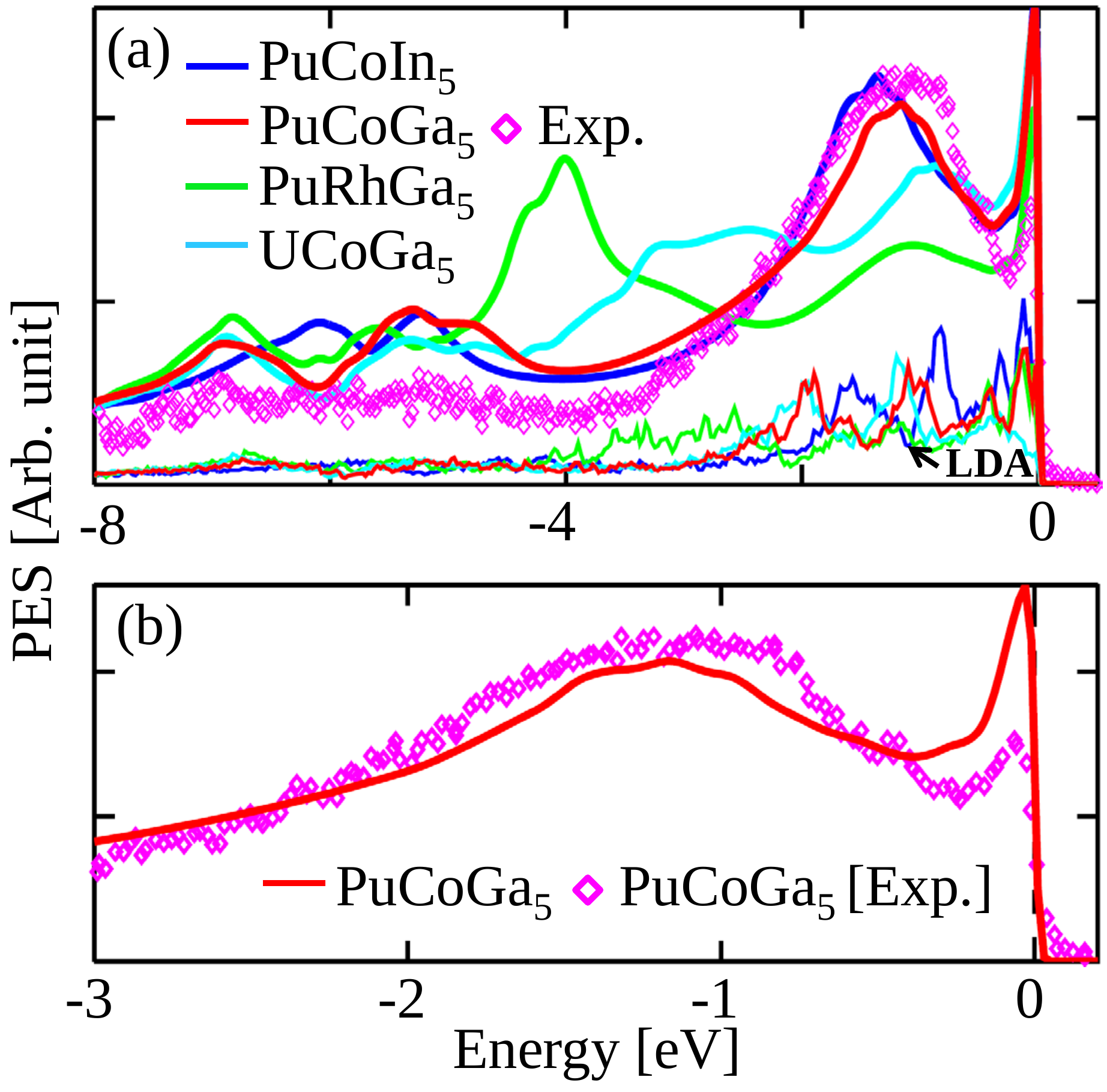}}
 \caption{(Color online) (a) Computed PES spectra for various compounds are compared with data for PuCoGa$_5$ \cite{JoycePRL03}. (b) Zoomed in view of (a) for PuCoGa$_5$ spectrum. All theoretical spectra have been renormalied by the same scaling factor.} \label{pes_1imp}
\end{figure}

At low energies,  when $\Sigma^{\prime}>0$, all quasiparticle states are renormalized toward $E_F$, see the quasiparticle spectra in Figs.~\ref{sig_1imp}(c)-(f). In this energy region, $\Sigma^{\prime\prime}$ is small, reflecting that quasiparticle states are coherent and itinerant. Above the peak in $\Sigma^{\prime\prime}$, where $\Sigma^{\prime}<0$, quasiparticle states are pushed to higher energy. The lost spectral weight from the peak in $\Sigma^{\prime\prime}$ is redistributed toward low energy near 1~eV in binding energy. A similar spectral weight redistribution occurs at the second peak (hump) in  $\Sigma^{\prime\prime}$ near 2~eV binding energy. As a result further pile-up of spectral weight occurs around
1.0-1.5 eV, creating new quasiparticle states due to electronic correlations. The quasiparticle states in this energy region are incoherent and fairly dispersionless, reflecting the dual aspect of the localized behavior of 5$f$ electrons. {Qualitatively similar behavior was also found by using the LDA+DMFT method, however, with a weaker renormalization toward the Fermi level \cite{Pourovskii}.}

To compare our calculations with experiment, we compute the PES spectra as $I_{PES}=\bigl<A({\bf k},\omega)\bigr>_kn_F(\omega)$ (neglecting any matrix-element effects). We compare with available data for PuCoGa$_5$ at 77~K \cite{JoycePRL03} shown by magenta diamonds in Fig.~\ref{pes_1imp}. Good quantitative agreement is evident. Near $E_F$ experiment shows a broader feature than theory with less spectral weight, which may be related to experimental resolution and theoretical approximations. The present calculation slightly underestimates the dip in the spectral weight, which stems from the neglect of orbital matrix-elements, charge and other fluctuations, as well as the quasiparticle approximation in the self-consistency scheme of the calculation of the self-energy. The key result is that both the spectral weight loss at low energy and high energy are well captured by the spin-fluctuation model. As we move across the series from PuCoIn$_5$ to UCoGa$_5$ the spectral weight redistribution gradually decreases. This suggests that spin-fluctuations play a lesser role in UCoGa$_5$ than in the isostructural Pu-115 compounds.

\begin{figure}[top]
\rotatebox[origin=c]{0}{\includegraphics[width=0.9\columnwidth]{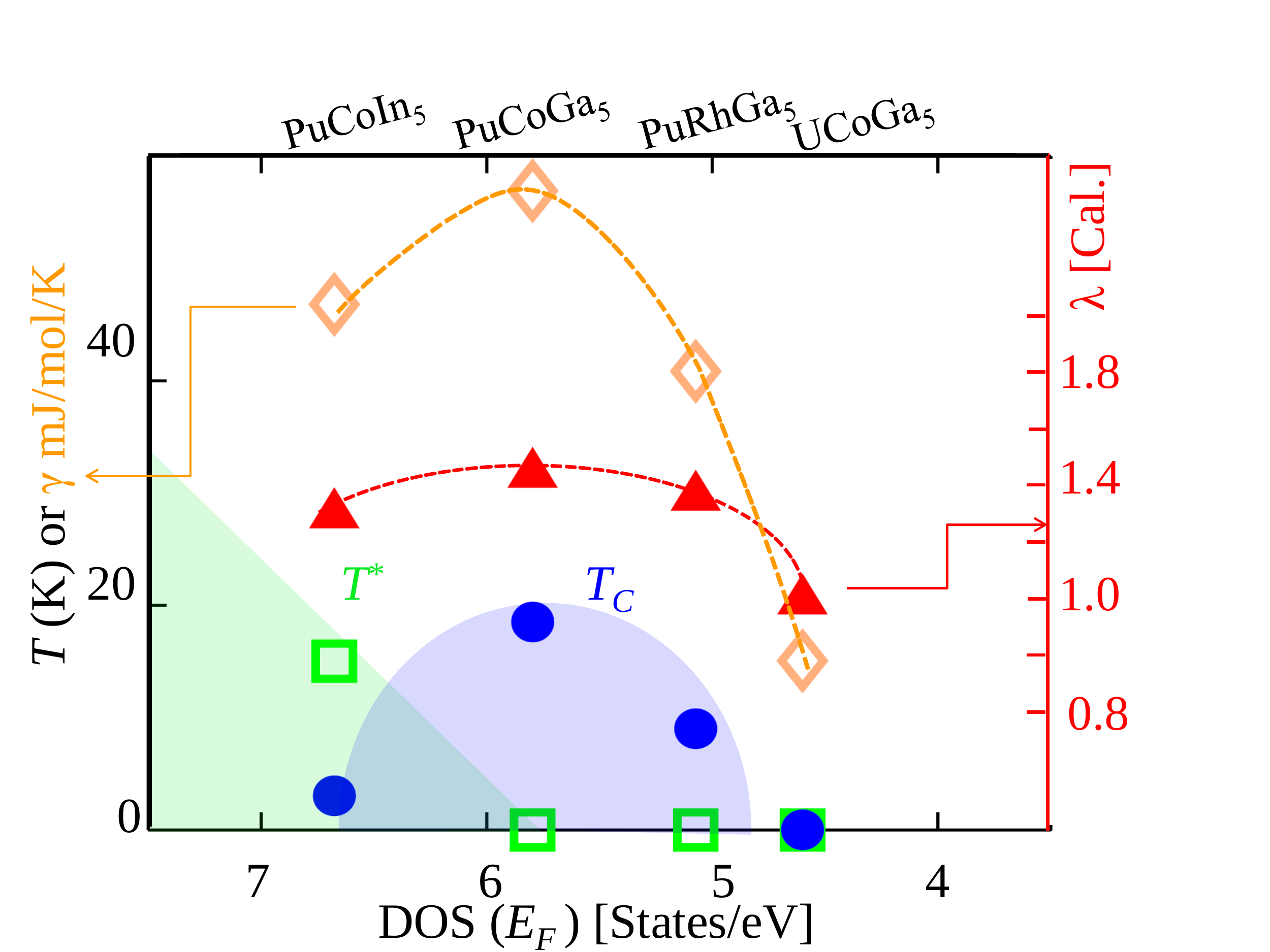}}
 \caption{{(Color online) Experimental values of $T_c$ and an impurity phase $T^*$ \cite{EDBauer:2011} are plotted as a function of  the bare DOS at $E_F$ (theory) and compared with computed values of the spin-fluctuation coupling constant $\lambda$ and corresponding Sommerfeld coefficient $\gamma$.}} \label{phase_1imp}
\end{figure}

Finally, we calculate the spin-fluctuation coupling constant $\lambda$ from the energy derivative of $\Sigma^{\prime}$. In the low-energy region, we obtain $\Sigma^{\prime}(\xi_{\bf k})\approx-\lambda\xi_{\bf k}=(1-Z^{-1})\xi_{\bf k}$. The coupling constant $\lambda$ follows the same material dependence as $T_c$ across the series from PuCoIn$_5$$\rightarrow$UCoGa$_5$ with its maximum for PuCoGa$_5$. Although $\lambda$ is quite large for PuCoIn$_5$, its $T_c$ is strongly suppressed probably due to competition with an impurity phase \cite{EDBauer:2011}. Our estimation of the fluctuation renormalized Sommenfeld coefficient $\gamma$ follows $T_c$ in Fig.~\ref{phase_1imp}. For PuCoGa$_5$, we find the renormalized $\gamma=57$~mJ/mol/K$^2$, which is slightly less than the corresponding experimental value of 77~mJ/mol/K$^2$ \cite{SarraoYEAR02}, suggesting room for phonon fluctuations of about $\lambda_{ep}\sim0.8$, which is very close to the electron-phonon coupling constant deduced by first-principles calculations \cite{phonon}. Note that our calculated coupling constant of $\lambda=1.4$ for PuCoGa$_5$ is smaller than the calculated value of 2.5 obtained within the LDA+DMFT approximation \cite{LDMA}.

In conclusion, we presented a first-principles based intermediate coupling model for calculating the multiband spin-fluctuation spectrum within the GW method. The presence of a strong spin-fluctuation peak in $\Sigma^{\prime\prime}$ is found around 0.5 eV, which splits the electronic states into an itinerant coherent part close to $E_F$ and strongly localized incoherent states around 1.0-1.5 eV. These results agree well with the experimental peak-dip-hump PES structure \cite{JoycePRL03}. In addition, the isostructural Pu-115 and UCoGa$_5$ compounds (for UCoGa$_5$ the 5$f$ electrons are moved above $E_F$)  have qualitatively similar electronic band structure near $E_F$. This can be understood approximately within a unified rigid-band filling scheme, which can account for band shifts through controlled hole doping. {Finally, we calculated a spin-fluctuation coupling constant $\lambda$ of order unity. It follows the same materials dependence as $T_c$, indicating that spin-fluctuation mediated pairing is a strong candidate for superconductivity in these materials.}

\begin{acknowledgments}
We thank A. V. Balatsky, E. D. Bauer, F. Ronning, T. Durakiewicz, and J. J. Joyce for discussions. We are especially grateful to E.D.B.\ and F.R.\ for sharing their unpublished data on PuCoIn$_5$. Work at the Los Alamos National Laboratory was supported by the U.S.\ DOE under Contract No.\ DE-AC52-06NA25396 through the Office of Science (BES) and the LDRD Program.
We acknowledge a NERSC computing allocation of the U.S.\ DOE under Contract No.\ DE-AC02-05CH11231.
\end{acknowledgments}

\section{Supplementary Information}
{\it Intermediate coupling model$-$}
We calculate materials-specific first-principles electronic band structures,
including spin-orbit coupling, within the framework of density functional theory in the generalized-gradient
approximation (GGA)~\cite{JPPerdew:1996}. We use
the full-potential linearized augmented plane wave method of WIEN2k. The calculation is performed with 40 bands to capture the $\pm$10 eV energy window of  relevance around the Fermi level. The spectral representation of the Green's function is constructed as
\begin{equation}
G_{sp}({\bf k},i\omega_n) = \sum_{\mu}\frac{\phi_{\mu}^s({\bf k})\phi_{\mu}^{p*}({\bf k})}{i\omega_n-E_{\mu}({\bf k})}.
\end{equation}
Here $\phi_{\mu}^i$ is the eigenstate for $\mu^{th}$ band $E_{\mu}$ projected on the $i^{th}$ orbital. The non-interacting susceptibility in the particle-hole channel represents joint density of states (JDOS), which can be calculated by convoluting the corresponding Green's function over the entire Brillouin zone (BZ) to obtain (spin and charge bare susceptibility are the same in the paramagnetic ground state)\cite{Graser}:
%
\begin{equation}\label{chibare}
\chi^0_{spqr}({\bf q},\Omega)=-\frac{T}{N}\sum_{{\bf k},n}G_{sp}({\bf k},i\omega_n)G_{qr}({\bf k+q},i\omega_n+\Omega).
\end{equation}
Within the RPA, spin and charge channels become decoupled. (We ignore particle-particle as well as  weaker charge fluctuation processes.) In the spin channel, the collective many-body corrections of the spin-fluctuation spectrum can be written in matrix representation:
$\hat{\chi}=\hat{\chi}^0[\hat{1}-\hat{U}_s\hat{\chi}^0]^{-1}$.
The interaction matrix $\hat{U}_s$ is defined in the same basis consisting
of intra-orbital $U$, inter-orbital $V$, Hund's couling $J$ and pair-scattering $J^{\prime}$ terms \cite{Graser,Dastworesonance,DasFeSe}. In the present calculation, we neglect the orbital overlap of eigenstates, i.e., we assume $\phi^i_{\mu}=1$ when $i=\mu$. Such an approximation simplifies the calculation and $\hat{\chi}^0$ becomes a diagonal matrix and $J=J^{\prime}=0$.

Using the GW approximation, where $G$ represents the Green's function and $W$ is the interaction
vertex, we write the spin-fluctuation interaction vertex  following Ref.~\cite{Bickers}:
%
\begin{equation}
V_{pqrs}({\bf q},\Omega) = \left[\frac{3}{2}\hat{U}_s \hat{\chi}^{\prime\prime}({\bf q},\Omega)\hat{U}_s\right]_{pqrs} .
\label{EQ:Vmatrix}
\end{equation}
The Feynmann-Dyson equation for the  imaginary part of the self-energy in a multiband system with $N$  sites is
%
\begin{eqnarray}
%
%
%
&&\Sigma_{pq}^{\prime\prime}({\bf k},\omega) = -\frac{1}{N}\sum_{\bf q}\sum_{rs}\int_{-\infty}^{\infty}d\Omega~\Gamma\, V_{pqrs}({\bf q},\Omega)\nonumber\\
&&~~~~\times\Bigl[ A_{rs}({\bf k}+{\bf q},\omega+\Omega)(n_B(\Omega)+n_F(\omega+\Omega))\nonumber\\
&&~~~~+A_{rs}({\bf k}+{\bf q},\omega-\Omega)(n_B(\Omega)+1-n_F(\omega-\Omega))\Bigr].\label{Eq:selfenergy}
\end{eqnarray}
%
%
%
The quasiparticle spectral weight is defined by $A_{rs}({\bf k},\epsilon)=-{\rm Im}\bigl[G_{rs}({\bf k},\epsilon)\bigr]/\pi$. The quantities $n_B$ and $n_F$ are the Bose-Einstein and Fermi-Dirac distribution functions, respectively. $\Gamma$ is the vertex correction discussed later. For a more accurate calculation, one
needs to account for the anisotropy in $\hat V({\bf q},\Omega)$.
In the present case, where the spin-fluctuation spectrum is only weakly anisotropic (see Fig.~2), it is justified to replace the first term in Eq.~(\ref{Eq:selfenergy}) by a momentum-averaged spin-fluctuation function interaction, that is,
$\bigl<\hat V(\Omega)\bigr>_{\bm{q}}=\left[\frac{a^3}{(2\pi)^3}\right]\int d^3q \hat V({\bf q},\Omega)$. 
This is equivalent to dropping the $\bm{k}$ dependence of the self-energy, which greatly simplifies the numerical self-consistency loop.
It then follows from Eq.~(\ref{Eq:selfenergy}) that  at $T=0$, the imaginary part of  $\Sigma_{pq}(\bm{k}, \omega)$ reduces to
\begin{equation}
\Sigma_{pq}^{\prime\prime}(\omega) \approx -2\sum_{rs}\int_{0}^{\omega} d\Omega\,
\Gamma \, \bigl<V_{pqrs}(\Omega)\bigr>_{\bm{q}} N_{rs}(\omega-\Omega),
%
\label{Eq:selfenergy2}
\end{equation}
for $\omega>0$, where the density of states is given by $N_{rs}(\epsilon)=\sum_{\bf k}A_{rs}({\bf k},\epsilon)$.
(For $\omega < 0$, the only changes are that the upper limit of the integral is $|\omega|$ and the argument of $N_{rs}$ is $\Omega-|\omega|$, which is $<0$.) We use Eq.~(\ref{Eq:selfenergy2})
to compute the imaginary part of the self-energy from the first-principles band structure. The real part of the self-energy, $\Sigma_{pq}^{\prime}(\omega)$, is obtained by using the Kramers-Kronig relationship. Finally, the self-energy dressed quasiparticle spectrum is determined by Dyson's equation:
\begin{equation}
\hat{G}^{-1}(\bm{k}, \omega)=\hat{G}_{0}^{-1}(\bm{k}, \omega)-\langle \hat{\Sigma}(\bm{k},\omega) \rangle_{\bm{k}} .
\end{equation}


\end{document}